\documentclass[%
 reprint,
 amsmath,amssymb,
 aps,
 prl,
]{revtex4-2}

\usepackage{graphicx}
\usepackage{dcolumn}
\usepackage{bm}
\usepackage[usenames]{color}
\usepackage{colortbl}

\begin{document}

\title{Terahertz frequency conversion at plasma-induced time boundary}

\author{Yindong Huang$^{1}$}
\email{yindonghuang@nudt.edu.cn}
\author{Bin Zhou$^{1,2}$, Aijun Xuan$^{1,3}$, Mingxin Gao$^{1,4}$, Jing Lou$^{1}$,  Xiaomin Qu$^{1}$,\\  Zengxiu Zhao$^{4}$}
\author{Ce Shang$^{5}$}
\email{shangce@aircas.ac.cn}
\author{Xuchen Wang$^{6}$}
\email{xuchen.wang@hrbeu.edu.cn}
\author{Chao Chang$^{1,7}$}
\email{gwyzlzssb@pku.edu.cn}
\author{Viktar Asadchy$^{8}$}

\affiliation{$^1$Innovation Laboratory of Terahertz Biophysics, National Innovation Institute of Defense Technology, Beijing, 100071, China}
\affiliation{$^2$Department of Physics, Tsinghua University, Beijing  100084, China}
\affiliation{$^3$School of Mathematics and Physics, North China Electric Power University, 102206 Beijing, China}
\affiliation{$^4$Department of Physics, National University of Defense Technology, Changsha 410073, China}
\affiliation{$^5$Aerospace Information Research Institute, Chinese Academy of Sciences, Beijing, 100094 China}
\affiliation{$^6$Qingdao Innovation and Development Base, Harbin Engineering University, Qingdao, China}
\affiliation{$^7$School of Physics, Peking University, Beijing, 100871, China}
\affiliation{$^8$Department of Electronics and Nanoengineering, Aalto University, Maarintie 8, 02150 Espoo, Finland}

\date{\today}
\begin{abstract}
We report on the frequency conversions of terahertz (THz) waves at ultrafast time boundaries created via femtosecond laser-induced air-to-plasma phase transitions. Our combined experimental and theoretical approach reveals that the abrupt change in refractive index at the ultrafast time boundaries drives both the red and blue shifts over the broadband THz spectrum due to the dispersive plasma, with distinctive amplitude variations. The present study contrasts these effects with those from spatial boundaries, highlighting the superior efficacy of temporal manipulations for spectral engineering. These findings not only deepen the understanding of light-matter interactions in time-varying media but also pave the way for innovative applications in THz technology and lay the groundwork for the observation of temporal reflection effects, photonic time crystals, and spatio-temporally modulated matter.
\end{abstract}

\maketitle
\begin{figure}[tbp]
\includegraphics[width=8 cm]{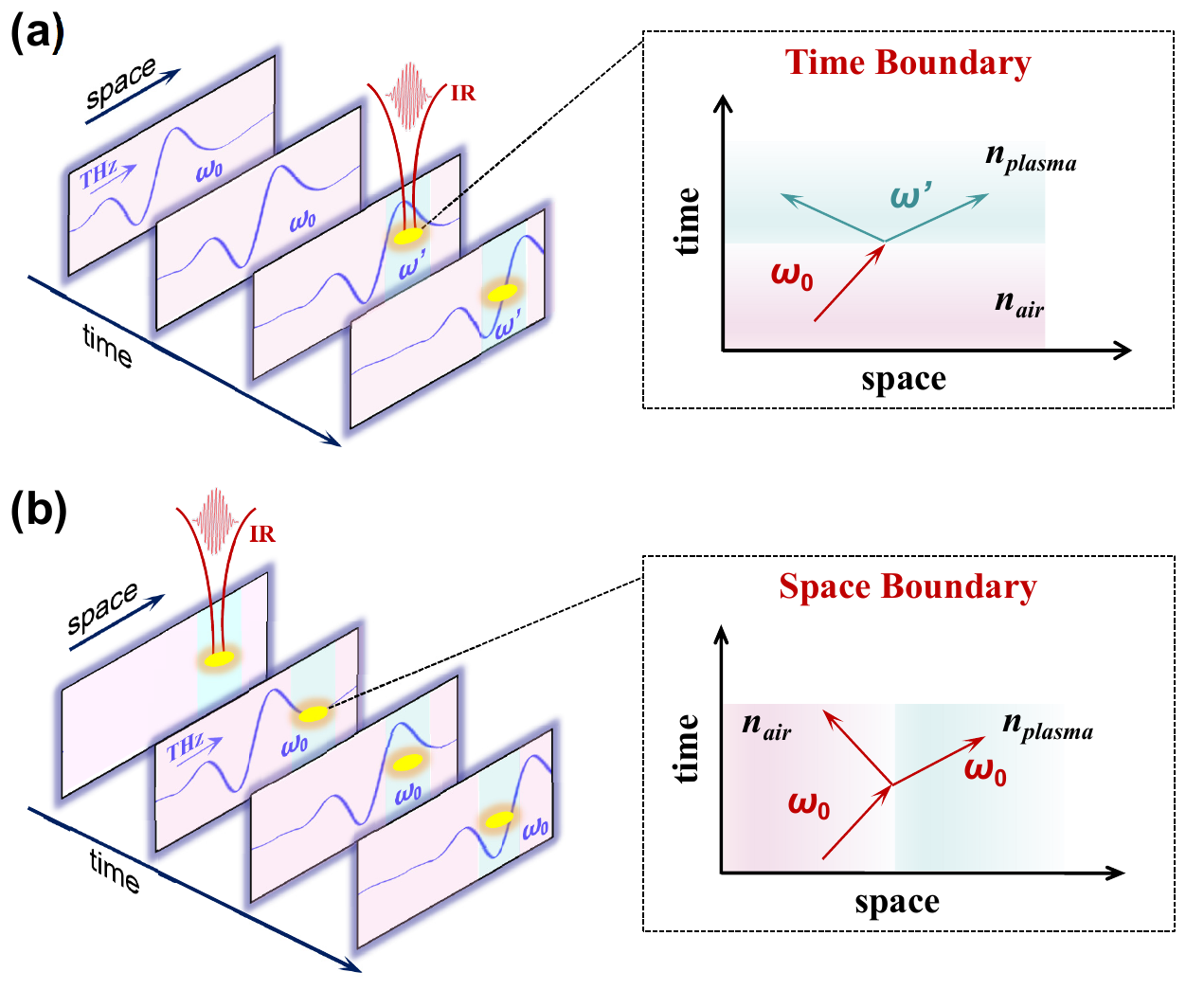}
\caption{Schematic diagram comparing the time- and space boundaries of the air-to-plasma phase transition induced by the laser pulse. (a) Plasma is generated during the propagation of the THz wave to induce an abrupt refractive index change in the time domain due to the laser ionization. The inset shows the time-reflection and time-refraction inducing the frequency shift of THz waves at the time boundary. (b) When plasma is generated before the arrival of the THz wave, it acts as a space boundary with no frequency shift.
}
	\label{f1}
\end{figure}

Fundamental states of matter—gas, liquid, solid, and plasma—originate from distinct atomic or molecular configurations and condensation dynamics. Phase transitions between these states induce variations in electromagnetic properties (e.g., refractive index) through structural reorganization. The refractive index governs electromagnetic-wave propagation via the reflection and refraction at the spatial boundaries with abrupt refractive index discontinuities. 
Unlike the spatial boundaries, a {\it time boundary} refers to an abrupt, transient change in refractive index over time rather than space (see FIG.~\ref{f1}), enabling exotic phenomena such as time-refraction and time-reflection, where light undergoes frequency shifts \cite{Plansinis2015},  total internal reflection \cite{Bar-Hillel2024}, and spatio-spectral fission without spatial interfaces \cite{Jaffray2025}.  
These dynamic light-matter interactions underpin the concept of photonic time crystals, achieved through periodic refractive index modulation in time \cite{Asgari2024,Hayran2025}, and unlock transformative applications including optical amplification \cite{Lyubarov2022time_crystal,Wang2025}, harmonic generation \cite{Konforty2025}, superlattice \cite{DongZH2025}, localization and scattering \cite{Sharabi2021,KimJM2023}, time grating \cite{Wood2020}, spatio-temporally engineered photonics \cite{Sharabi2022Optica,Wang2023} and others.

Recent experimental advances have witnessed spectral control at time boundary across multiple frequency regimes. In the microwave range, time-reflection has been achieved using electrically modulated metamaterials \cite{Moussa2023} and optically controlled photo-diodes \cite{Jones2024}. In the near-infrared, time-refraction induced by photoexcitation of transparent conducting oxides (TCOs) \cite{Alam2016} has enabled time-slit interference \cite{Tirole2023}, spectral shifts in fundamental waves \cite{Zhou2020, Bohn2021, Eran2023}, and terahertz (THz) generation \cite{Lu2025}.
However, establishing ultrafast time boundaries at high-frequencies remains fundamentally challenging, as it demands substantial modulation depth of the refractive index on timescales comparable to the optical oscillation period.
Conventional approaches based on nonlinear optical materials and TCOs have thus far demonstrated limited modulation speed and depth, highlighting the need for alternative strategies to achieve optical time-reflection and photonic time crystals.

To overcome the above challenges, plasma-based approaches emerge as a compelling alternative.
Femtosecond laser-induced phase transitions from neutral gas to plasma achieve refractive index contrasts of approximately 100\% or more within tens of femtoseconds, enabling sub-cycle temporal control of the refractive index for THz and higher-frequency waves. This substantial contrast, determined by plasma dispersion, can be tuned by varying the plasma density \cite{zheng2017filament}.
Unlike conventional materials such as TCOs, plasmas do not require operation near the epsilon-near-zero regime to achieve strong permittivity contrast modulation.
These attributes—ultrafast transition speeds, substantial refractive index contrast, and tunable frequency range—position laser-induced plasma as a powerful platform for THz photonics and the implementation of time boundaries.

In this Letter, we demonstrate THz frequency shifting via time-refraction, driven by femtosecond laser-induced air-to-plasma phase transitions.
Within the 100 fs femtosecond laser pulse duration, abrupt refractive index changes at the air-to-plasma time boundaries induce both red and blue shifts across the broadband THz spectrum, with distinctive amplitude variations.
By integrating the tunneling ionization model for plasma preparation with the THz-band refractive index of plasma, we provide theoretical validation of the experimental results. Through precise control of the time delay between plasma formation and THz arrival, we observe the time-dependent spectral reshaping and confirm time-refraction as an intrinsic feature of ultrafast time boundaries. These findings establish phase transitions as a practical toolset for plasma-based optics and the engineering of dynamic photonic materials.

\begin{figure*}[tbp]
\includegraphics[width=17 cm]{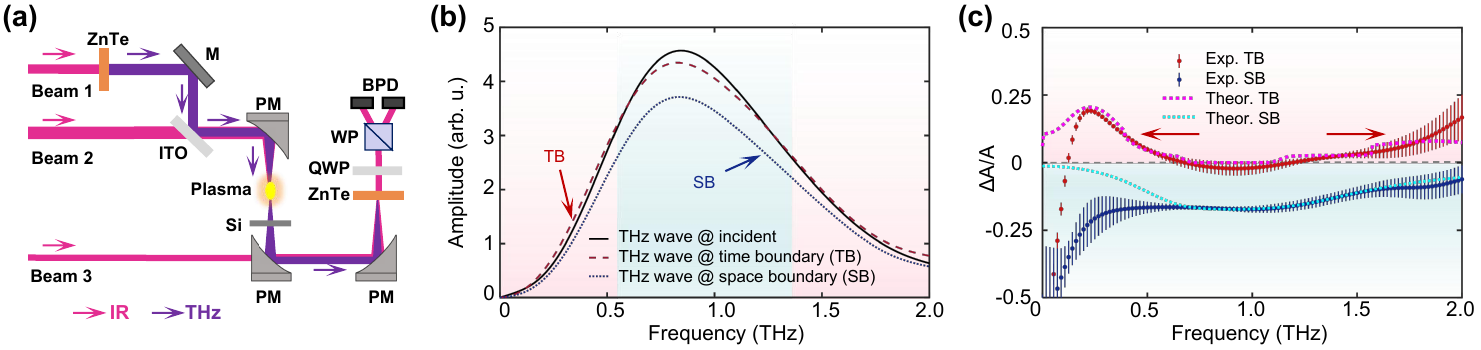}
\caption{(a) Experimental setup. ZnTe: zinc telluride crystal. M: mirror. ITO: indium tin oxide glass. PM: parabolic mirror. Si: silicon plate. QWP: quarter wavelength plate. WP: Wollaston prism. BPD: balanced photo-diodes.
(b) The Fourier-transformed THz amplitudes of the incident THz waves (black solid lines), transmitted THz waves with a time boundary (red dashed lines), and transmitted THz waves with a space boundary (blue dotted lines).
(c) The relative THz amplitude variation ($\Delta A/A$) versus the THz frequencies for the time boundary (red points) and space boundary (blue points) with error bars, as well as the theoretical calculation results of the time-boundary (TB, pink dotted lines) and the space boundary (SB, light-blue dotted lines). The two red arrows indicate the bidirectional frequency shift induced by the time boundary.}
	\label{f2}
\end{figure*}

Experimentally, we split a femtosecond infrared laser (100 fs, 800 nm, 5.4 mJ, 1 kHz) into three beams for THz wave generation, plasma formation, and electro-optic sampling, respectively [FIG.~\ref{f2}(a)]. Beam 1 passed through a 1-mm-thick (110)-cut zinc telluride (ZnTe) crystal to produce THz waves, which then sequentially reflected by a mirror and a 0.5-mm-thick indium tin oxide (ITO) glass. High-power Beam 2 co-propagated with the reflected THz waves and ionized the gaseous molecules at the focal point of an off-axis parabolic mirror with a 7.62-cm focal length. A high-resistance silicon wafer was placed after the focal point, blocking the femtosecond laser while transmitting the THz wave. Beam 3 was focused 
onto a 1-mm-thick zinc telluride (ZnTe) crystal along with the interacting THz wave for electro-optic sampling of the transmitted THz waveform. By adjusting the time delay $t$ between Beam 1 and Beam 2, we could switch between forming time- and space-boundaries.

As shown in FIG.~\ref{f1}(a), synchronized femtosecond laser ionization establishes a transient temporal plasma interface during THz wave propagation, inducing measurable spectral shifts via time-refraction.
The inset illustrates frequency conversion from $\omega_0$ to $\omega'$ across this time boundary. The abrupt change in the real part of refractive index $n_{\rm{p,r}}(\omega',t)$ alters the phase velocity of light while conserving momentum, thus resulting in the frequency shifts according to the relation:
\begin{equation}
  \omega^{\prime}\cdot {n_{\text {p,r}}(\omega',t)}=\omega_0 \cdot n_{\text {air}},  
\end{equation}
where red shifts occur when $n_{\text {p,r}}(\omega',t)>n_{\text {air}}$ and blue shifts when $n_{\text {p,r}}(\omega',t)<n_{\text {air}}$. Here, the subscript r denotes the real part. In contrast to the space boundaries [Fig.~\ref{f1}(b)], these dynamically formed time boundaries uniquely enable frequency conversion—a process dictated by the relative time delay between plasma generation and THz wave interaction. 

We measured the transmission spectra of THz waves through laser-induced plasma (1 mJ laser energy input) at varying time delay $t$ to compare time- and space-boundaries. As illustrated in FIG.~\ref{f2}(b), the black lines represent the Fourier-transformed THz spectrum as a reference. The time-boundary interaction (red dashed lines) exhibits a dual-band response with gain enhancement (pink region) and suppression (cyan region), indicating a frequency-selective modulation that contrasts sharply with our previous observation of broadband amplification or suppression using two-color laser-prepared plasma dipoles \cite{Qu2024}. This highlights a distinct mechanism of the monochromatic fundamental field-generated plasma in the present work.
Furthermore, upon the arrival of the THz wave in the interaction region subsequent to plasma generation, i.e., space boundary, the transmission spectrum (blue dotted lines) exhibits pronounced broadband attenuation across the entire THz band. This spatial interface behavior aligns with the previously reported results of confinement, absorption, and local-field enhancement of the input THz waves at the air-plasma spatial interfaces \cite{WangXK2022, ZhaoJY2023, HuangYD2021PRA,zheng2017filament}.

To distinguish the time- and space-boundary effects quantitatively, we define a parameter for the relative THz amplitude modulation, $\Delta A/A = (A_{\rm post}- A_{\rm pre})/A_{\rm pre}$, where $A_{\rm pre}(\omega)$ and $A_{\rm post}(\omega)$ denote the THz amplitudes before and after interaction with the plasma at frequency $\omega$.
Positive values of $\Delta A/A$ indicate amplitude enhancement, while negative values indicate attenuation.
In FIG.~\ref{f2}(c), the color-coded regions directly show this frequency-selective enhancement and attenuation. Under time-boundary conditions, THz amplitude enhancements occur at both ends of the measured THz spectra, with maximum gains of $\sim$19.2\% at 0.2 THz in the low-frequency wing and $\sim$16.7\% at 2.0 THz in the high-frequency wing. 
This dual-band response suggests the bidirectional frequency shifting, i.e., concurrent blue- and red-shifted  THz spectrum components, in contrast with the unidirectional shifts reported in the infrared regime \cite{Eran2023}. The strongest attenuation of $\sim$2.2\% occurs at around 0.9 THz in the middle frequency range, indicating a unique time-boundary effect in the air-to-plasma phase transition.
In contrast, space-boundary interaction causes 10-20\% attenuation across the entire 0.3-2.0 THz spectral range, clearly distinguishing the two regimes.

\begin{figure*}[tbp]
\includegraphics[width=17 cm]{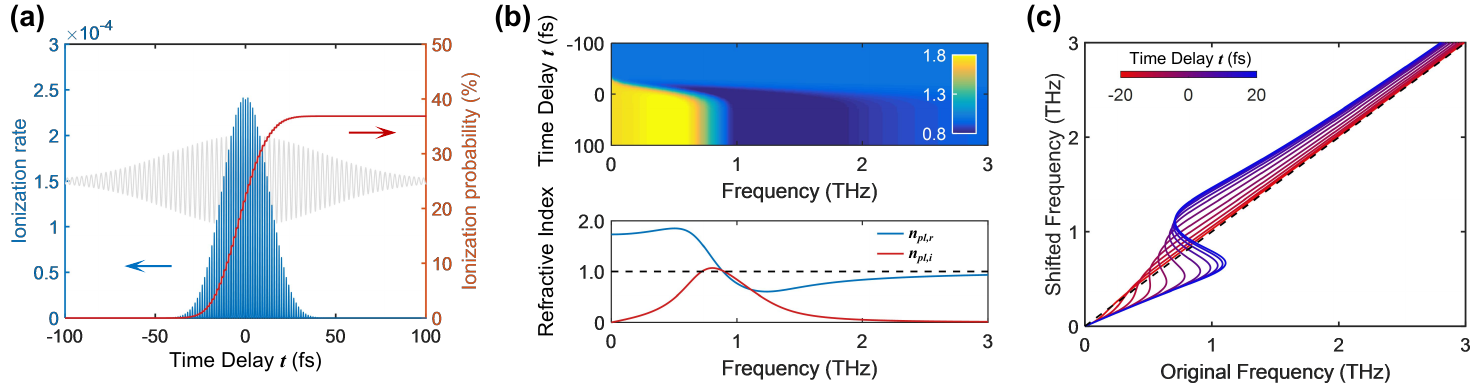}
\caption{(a) Theoretically calculated tunneling ionization rate (left y-axis, blue solid lines) and the ionization probability (right y-axis, red solid lines) using the MO-ADK model \cite{TongXM2002}. The gray lines indicate the shape of the incident laser fields.
(b) The upper panel shows the time- and frequency-dependent real part of the refractive index. In the lower panel, the real parts (blue solid lines) and the imaginary parts (red solid lines) of the refractive index are compared with time delay $t = 100$ fs.
(c) Time delay $t$ dependent frequency variation from the original frequency to the shifted frequency. The zero time delay is set at the peak of the envelope of the laser pulse. The dashed line indicates cases where the original frequency does not shift. The region below the dashed line corresponds to red shift, while the region above corresponds to blue shift.}
	\label{f3}
\end{figure*}

To elucidate the bidirectional frequency shifting observed at the ultrafast time boundary, we developed a theoretical framework that integrates the plasma formation dynamics and time refraction. When exposed to an intense femtosecond laser, gaseous molecules undergo tunneling ionization \cite{Smirnov1966, Ammosov1986}. As illustrated in FIG.~\ref{f3}(a), by employing the molecular Ammosov-Delone-Krainov (MO-ADK) model with the ground-state depletion \cite{TongXM2002}, we calculated the tunneling ionization rates and cumulative probabilities of N$_2$ molecules under a Gaussian pulse (100 fs FWHM) with a peak intensity $\sim$1.7$\times$10$^{14}$~W/cm$^2$.
Tunneling ionization predominately occurs in bursts synchronized with the laser field's sub-cycle peaks, producing a step-like electron density $\rho_e(t)$ within 100 fs, shorter than the ps-scaled THz oscillating period. This creates an ultrafast time boundary for the propagating THz waves. Note that the time delay $t$ is defined within the pulse envelope, with $t=0$ corresponding to the envelope peak.

The step-like air-to-plasma phase transition induces an abrupt, time-varying change in the refractive index experienced by the transmitted THz waves. Assuming that the laser-ionized plasma forms a cylinder with uniform electron density $\rho_e(t)$, the external THz field generates a transient local field within the plasma \cite{mics2005nonresonant,zheng2017filament}. This local field resonates with the incident THz waves, thereby modifying the refractive index in the THz frequency band. Given the picosecond timescale of the time-boundary interaction, we attribute the electron density solely to tunneling ionization and neglect any further evolution from collisional ionization or electron–particle recombination during the THz–plasma interaction. Under these assumptions, the laser-induced plasma behaves as a dispersive medium with a time-varying refractive index. The transient complex plasma dielectric response during laser excitation can be expressed as:
\begin{equation}
\varepsilon_{\rm p} (\omega, t) = 1 - \frac{\omega_{\rm p}(t)^2}{\omega^2 - \omega_{\rm p}(t)^2/2 + i \gamma\omega},
\end{equation}
where $\omega_{\rm p}(t) = \sqrt{ \rho_e(t) e^2/ \varepsilon_0 m_e}$ is the time-dependent plasma frequency and $\gamma$ is the collisional damping rate \cite{mics2005nonresonant,zheng2017filament}. The term $\omega_{\rm p}^2/2$ arises from the screening effect of electrons in the plasma, as detailed in Sec. 1 of the Supplemental Material (SM).
This formalism allows calculation of the time-dependent plasma refractive index $n_{\rm p,r}(\omega, t) = \mathfrak{Re} [\sqrt{\varepsilon_{\rm p}(\omega,t)}]$.
As shown in the upper panel of FIG.~\ref{f3}(b), the refractive index equals unity across all frequencies for $t < -20$ fs. At $t = 0$, the real part of the refractive index becomes dispersive and deviates from unity, consistent with the ionization sketched in FIG.~\ref{f3}(a), where ionization onset occurs at approximately $-20$ fs. After $t = 40$ fs, the ionization probability saturates, and the electron density, along with the refractive index, becomes time-independent.

Within the laser pulse duration, the propagating THz waves undergo frequency shifts from $\omega_0$ to $\omega'$, governed by Eq.~(1). This is attributed to the significant changes in the refractive index of the dispersive plasma near the resonant frequency, which permits multiple shifted frequencies that satisfy the momentum-conservation condition at the time boundary \cite{Solis2021}.
As shown in FIG.~\ref{f3}(c), as the plasma density increases, the refractive index variations can cause the original frequency to convert into multiple frequencies, producing a bistable frequency variation curve. For instance, the blue curve at $t=20$ fs in FIG.~\ref{f3}(c) exhibits an ``S''-shaped bistable structure. The 
amplitude of the shifted frequency (vertical axis) behavior can be categorized into two regions: at the spectral ends, a single original frequency converts into a single shifted frequency, observed in the low-frequency (0-0.4 THz) and high-frequency ($>$1.5 THz) THz band; in the mid-frequency THz band (0.4-1.5 THz), a single original frequency can correspond to three shifted frequencies. For single-shifted-frequency cases, amplitude variation can be determined using time-refraction coefficient calculations \cite{Galiffi2022}. For the ``S''-shaped bistable structure with multiple shifted frequencies, refraction coefficients for two shifted frequencies at the time boundary are employed \cite{Solis2021}. 
In FIG.~\ref{f2}(c), experimental and theoretical amplitude changes under the time-boundary conditions are compared. Here, we only consider the highest and lowest shifted frequency points from the original frequency to the three converted frequency points, neglecting the non-steady frequency solutions (see Sec. 2 of the SM for details). This treatment has been performed in the lossless condition, and can be applied to calculate the transmission coefficients of time refraction~\cite{Solis2021}.  By separately calculating the amplitude variation for the single- and multiple-shifted frequency cases, theoretical results quantitatively reproduce the observed relative amplitude variations $\Delta A/A$.

\begin{figure}[tbp]
\includegraphics[width=8.5 cm]{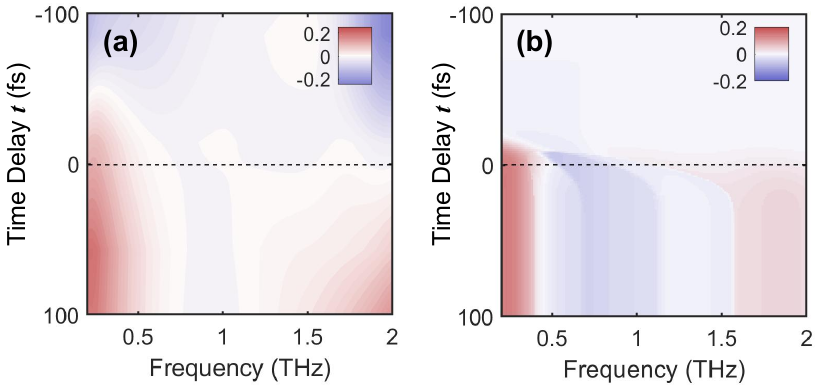}
\caption{Time delay $t$-dependent evolution of the relative THz amplitude variations ($\Delta A/A$) for the experimental (a) and theoretical results (b).}
	\label{f4}
\end{figure}

Based on the proposed method, the time delay $t$-dependent amplitude variations caused by different time boundaries can be calculated during the build-up process of electron density. Figure~\ref{f4}(a) presents the experimental results of THz amplitude variations, while FIG.~\ref{f4}(b) shows the theoretical calculations. These results exhibit qualitative agreement, though the specific values differ. Notably, the enhancement of low-frequency THz waves occurs significantly earlier than the high-frequency components. Additionally, both experiment and theory show attenuation of THz amplitudes at $t > 0$ near the plasma‐resonant frequency, corresponding to the unstable branch of the bistable frequency conversion.
Thus, the frequency conversion of THz waves has been effectively demonstrated at the time boundary of air-to-plasma phase transitions.

We next examine the space boundary effects on THz amplitude by considering the interface reflection and plasma absorption (detailed in Sec. 3 of the SM). For frequencies near the 0.9 THz resonance frequency (primarily 0.5-2.0 THz), the space-boundary-induced relative amplitude variations are quantitatively reproduced in FIG.~\ref{f2}(c). The calculations show good consistency with experiments, while the discrepancies between theory and experiment in the low-frequency range likely stem from the limited THz generation and electro-optic sampling efficiency. Moreover, since these low‐frequency components lie well below the resonance frequency, the refractive‐index resonance model becomes invalid.

Time-varying media with tunable refractive index modulation present emerging frontiers in photonic control. The present work establishes laser-induced air-to-plasma phase transitions as a promising approach for dynamic refractive index engineering. These ultrafast phase transitions enable precise spectral manipulation of electromagnetic waves, offering a novel paradigm for time-varying photonics. Resonance-enhanced refractive index contrast facilitates efficient frequency shifting, which is challenging to achieve in off-resonance conditions. Further efforts could explore the use of few-cycle or attosecond pulses to create more rapid time boundaries for near-infrared or visible light, and increasing the plasma volume to reveal time-reflection effects not observed in this study.
This temporal modulation approach provides additional opportunities for air-based electromagnetic wave modulation \cite{Schrodel2023AcoustoopticMO}. By tuning the plasma frequency via ionization, air—a ubiquitous, readily available medium—can be adapted to cover a broadband frequency range. 
Reconfigurable modulation of the spatial and temporal properties of light allows for the modulation of the transient photonic structure of the plasma with exceptional flexibility \cite{ZhangXB2025,Lehmann2016}.

In conclusion, we experimentally and theoretically demonstrate that laser-induced air-to-plasma phase transitions can generate ultrafast time boundaries with transient modulations of refractive index, inducing bidirectional frequency shifts and amplitude modulations in incident THz waves.
Using the MO-ADK model and incorporating refractive index variations, our theoretical calculations accurately reproduce the spectral changes induced by time boundaries at most frequencies. These changes contrast with the monotonous broad-spectrum attenuation observed under spatial boundaries. Our findings deepen the understanding of light-matter interactions at extreme time gradients and lay the groundwork for photonic time crystals and spatio-temporal mixing modulation.
Moreover, this phase-transition-based approach, analogous to hollow-fiber self-compression \cite{Wagner2004} but distinguished by its precise temporal and spatial control, may enable more efficient frequency conversion, pulse compression, soliton formation, and high-energy electron generation in future ultrafast photonic systems \cite{Pan2023,Jang2025,Dong2025,Ying2025}.

{\it Acknowledgment.} This work has been supported by the National Natural Science Foundation of China (Nos.~12174449, 12225511, T2241002), the Fundamental Research Funds for the Central Universities, China (No. 3072024WD2603), Project Nos. E4BA270100, E4Z127010F, E4Z6270100, E53327020D of the Chinese Academy of Sciences, and the Finnish Foundation for Technology Promotion. Yindong Huang and Bin Zhou contributed equally to this work.


\begin{thebibliography}{41}%
\makeatletter
\providecommand \@ifxundefined [1]{%
 \@ifx{#1\undefined}
}%
\providecommand \@ifnum [1]{%
 \ifnum #1\expandafter \@firstoftwo
 \else \expandafter \@secondoftwo
 \fi
}%
\providecommand \@ifx [1]{%
 \ifx #1\expandafter \@firstoftwo
 \else \expandafter \@secondoftwo
 \fi
}%
\providecommand \natexlab [1]{#1}%
\providecommand \enquote  [1]{``#1''}%
\providecommand \bibnamefont  [1]{#1}%
\providecommand \bibfnamefont [1]{#1}%
\providecommand \citenamefont [1]{#1}%
\providecommand \href@noop [0]{\@secondoftwo}%
\providecommand \href [0]{\begingroup \@sanitize@url \@href}%
\providecommand \@href[1]{\@@startlink{#1}\@@href}%
\providecommand \@@href[1]{\endgroup#1\@@endlink}%
\providecommand \@sanitize@url [0]{\catcode `\\12\catcode `\$12\catcode `\&12\catcode `\#12\catcode `\^12\catcode `\_12\catcode `\%12\relax}%
\providecommand \@@startlink[1]{}%
\providecommand \@@endlink[0]{}%
\providecommand \url  [0]{\begingroup\@sanitize@url \@url }%
\providecommand \@url [1]{\endgroup\@href {#1}{\urlprefix }}%
\providecommand \urlprefix  [0]{URL }%
\providecommand \Eprint [0]{\href }%
\providecommand \doibase [0]{https://doi.org/}%
\providecommand \selectlanguage [0]{\@gobble}%
\providecommand \bibinfo  [0]{\@secondoftwo}%
\providecommand \bibfield  [0]{\@secondoftwo}%
\providecommand \translation [1]{[#1]}%
\providecommand \BibitemOpen [0]{}%
\providecommand \bibitemStop [0]{}%
\providecommand \bibitemNoStop [0]{.\EOS\space}%
\providecommand \EOS [0]{\spacefactor3000\relax}%
\providecommand \BibitemShut  [1]{\csname bibitem#1\endcsname}%
\let\auto@bib@innerbib\@empty
\bibitem [{\citenamefont {Plansinis}\ \emph {et~al.}(2015)\citenamefont {Plansinis}, \citenamefont {Donaldson},\ and\ \citenamefont {Agrawal}}]{Plansinis2015}%
  \BibitemOpen
  \bibfield  {author} {\bibinfo {author} {\bibfnamefont {B.~W.}\ \bibnamefont {Plansinis}}, \bibinfo {author} {\bibfnamefont {W.~R.}\ \bibnamefont {Donaldson}},\ and\ \bibinfo {author} {\bibfnamefont {G.~P.}\ \bibnamefont {Agrawal}},\ }\bibfield  {title} {\bibinfo {title} {What is the temporal analog of reflection and refraction of optical beams?},\ }\href {https://doi.org/10.1103/PhysRevLett.115.183901} {\bibfield  {journal} {\bibinfo  {journal} {Phys. Rev. Lett.}\ }\textbf {\bibinfo {volume} {115}},\ \bibinfo {pages} {183901} (\bibinfo {year} {2015})}\BibitemShut {NoStop}%
\bibitem [{\citenamefont {Bar-Hillel}\ \emph {et~al.}(2024)\citenamefont {Bar-Hillel}, \citenamefont {Dikopoltsev}, \citenamefont {Kam}, \citenamefont {Sharabi}, \citenamefont {Segal}, \citenamefont {Lustig},\ and\ \citenamefont {Segev}}]{Bar-Hillel2024}%
  \BibitemOpen
  \bibfield  {author} {\bibinfo {author} {\bibfnamefont {L.}~\bibnamefont {Bar-Hillel}}, \bibinfo {author} {\bibfnamefont {A.}~\bibnamefont {Dikopoltsev}}, \bibinfo {author} {\bibfnamefont {A.}~\bibnamefont {Kam}}, \bibinfo {author} {\bibfnamefont {Y.}~\bibnamefont {Sharabi}}, \bibinfo {author} {\bibfnamefont {O.}~\bibnamefont {Segal}}, \bibinfo {author} {\bibfnamefont {E.}~\bibnamefont {Lustig}},\ and\ \bibinfo {author} {\bibfnamefont {M.}~\bibnamefont {Segev}},\ }\bibfield  {title} {\bibinfo {title} {Time refraction and time reflection above critical angle for total internal reflection},\ }\href {https://doi.org/10.1103/PhysRevLett.132.263802} {\bibfield  {journal} {\bibinfo  {journal} {Phys. Rev. Lett.}\ }\textbf {\bibinfo {volume} {132}},\ \bibinfo {pages} {263802} (\bibinfo {year} {2024})}\BibitemShut {NoStop}%
\bibitem [{\citenamefont {Jaffray}\ \emph {et~al.}(2025)\citenamefont {Jaffray}, \citenamefont {Stengel}, \citenamefont {Biancalana}, \citenamefont {Fruhling}, \citenamefont {Ozlu}, \citenamefont {Scalora}, \citenamefont {Boltasseva}, \citenamefont {Shalaev},\ and\ \citenamefont {Ferrera}}]{Jaffray2025}%
  \BibitemOpen
  \bibfield  {author} {\bibinfo {author} {\bibfnamefont {W.}~\bibnamefont {Jaffray}}, \bibinfo {author} {\bibfnamefont {S.}~\bibnamefont {Stengel}}, \bibinfo {author} {\bibfnamefont {F.}~\bibnamefont {Biancalana}}, \bibinfo {author} {\bibfnamefont {C.}~\bibnamefont {Fruhling}}, \bibinfo {author} {\bibfnamefont {M.}~\bibnamefont {Ozlu}}, \bibinfo {author} {\bibfnamefont {M.}~\bibnamefont {Scalora}}, \bibinfo {author} {\bibfnamefont {A.}~\bibnamefont {Boltasseva}}, \bibinfo {author} {\bibfnamefont {V.}~\bibnamefont {Shalaev}},\ and\ \bibinfo {author} {\bibfnamefont {M.}~\bibnamefont {Ferrera}},\ }\bibfield  {title} {\bibinfo {title} {Spatio-spectral optical fission in time-varying subwavelength layers},\ }\href {https://doi.org/10.1038/s41566-025-01640-1} {\bibfield  {journal} {\bibinfo  {journal} {Nat. Photonics}\ }\textbf {\bibinfo {volume} {19}},\ \bibinfo {pages} {558} (\bibinfo {year} {2025})}\BibitemShut {NoStop}%
\bibitem [{\citenamefont {Asgari}\ \emph {et~al.}(2024)\citenamefont {Asgari}, \citenamefont {Garg}, \citenamefont {Wang}, \citenamefont {Mirmoosa}, \citenamefont {Rockstuhl},\ and\ \citenamefont {Asadchy}}]{Asgari2024}%
  \BibitemOpen
  \bibfield  {author} {\bibinfo {author} {\bibfnamefont {M.~M.}\ \bibnamefont {Asgari}}, \bibinfo {author} {\bibfnamefont {P.}~\bibnamefont {Garg}}, \bibinfo {author} {\bibfnamefont {X.}~\bibnamefont {Wang}}, \bibinfo {author} {\bibfnamefont {M.~S.}\ \bibnamefont {Mirmoosa}}, \bibinfo {author} {\bibfnamefont {C.}~\bibnamefont {Rockstuhl}},\ and\ \bibinfo {author} {\bibfnamefont {V.}~\bibnamefont {Asadchy}},\ }\bibfield  {title} {\bibinfo {title} {{Theory and applications of photonic time crystals: a tutorial}},\ }\href {https://doi.org/10.1364/aop.525163} {\bibfield  {journal} {\bibinfo  {journal} {Adv. Opt. Photonics}\ }\textbf {\bibinfo {volume} {16}},\ \bibinfo {pages} {958} (\bibinfo {year} {2024})}\BibitemShut {NoStop}%
\bibitem [{\citenamefont {Hayran}\ and\ \citenamefont {Monticone}(2025)}]{Hayran2025}%
  \BibitemOpen
  \bibfield  {author} {\bibinfo {author} {\bibfnamefont {Z.}~\bibnamefont {Hayran}}\ and\ \bibinfo {author} {\bibfnamefont {F.}~\bibnamefont {Monticone}},\ }\bibfield  {title} {\bibinfo {title} {{A resonant tone for photonic time crystals}},\ }\href {https://doi.org/10.1364/aop.525163} {\bibfield  {journal} {\bibinfo  {journal} {Nat. Photonics}\ }\textbf {\bibinfo {volume} {19}},\ \bibinfo {pages} {126} (\bibinfo {year} {2025})}\BibitemShut {NoStop}%
\bibitem [{\citenamefont {Lyubarov}\ \emph {et~al.}(2022)\citenamefont {Lyubarov}, \citenamefont {Lumer}, \citenamefont {Dikopoltsev}, \citenamefont {Lustig}, \citenamefont {Sharabi},\ and\ \citenamefont {Segev}}]{Lyubarov2022time_crystal}%
  \BibitemOpen
  \bibfield  {author} {\bibinfo {author} {\bibfnamefont {M.}~\bibnamefont {Lyubarov}}, \bibinfo {author} {\bibfnamefont {Y.}~\bibnamefont {Lumer}}, \bibinfo {author} {\bibfnamefont {A.}~\bibnamefont {Dikopoltsev}}, \bibinfo {author} {\bibfnamefont {E.}~\bibnamefont {Lustig}}, \bibinfo {author} {\bibfnamefont {Y.}~\bibnamefont {Sharabi}},\ and\ \bibinfo {author} {\bibfnamefont {M.}~\bibnamefont {Segev}},\ }\bibfield  {title} {\bibinfo {title} {Amplified emission and lasing in photonic time crystals},\ }\href {https://doi.org/10.1126/science.abo3324} {\bibfield  {journal} {\bibinfo  {journal} {Science}\ }\textbf {\bibinfo {volume} {377}},\ \bibinfo {pages} {425} (\bibinfo {year} {2022})}\BibitemShut {NoStop}%
\bibitem [{\citenamefont {Wang}\ \emph {et~al.}(2025)\citenamefont {Wang}, \citenamefont {Garg}, \citenamefont {Mirmoosa}, \citenamefont {Lamprianidis}, \citenamefont {Rockstuhl},\ and\ \citenamefont {Asadchy}}]{Wang2025}%
  \BibitemOpen
  \bibfield  {author} {\bibinfo {author} {\bibfnamefont {X.}~\bibnamefont {Wang}}, \bibinfo {author} {\bibfnamefont {P.}~\bibnamefont {Garg}}, \bibinfo {author} {\bibfnamefont {M.~S.}\ \bibnamefont {Mirmoosa}}, \bibinfo {author} {\bibfnamefont {A.~G.}\ \bibnamefont {Lamprianidis}}, \bibinfo {author} {\bibfnamefont {C.}~\bibnamefont {Rockstuhl}},\ and\ \bibinfo {author} {\bibfnamefont {V.~S.}\ \bibnamefont {Asadchy}},\ }\bibfield  {title} {\bibinfo {title} {{Expanding momentum bandgaps in photonic time crystals through resonances}},\ }\href {https://doi.org/10.1038/s41566-024-01563-3} {\bibfield  {journal} {\bibinfo  {journal} {Nat. Photonics}\ }\textbf {\bibinfo {volume} {19}},\ \bibinfo {pages} {149} (\bibinfo {year} {2025})}\BibitemShut {NoStop}%
\bibitem [{\citenamefont {Konforty}\ \emph {et~al.}(2025)\citenamefont {Konforty}, \citenamefont {Cohen}, \citenamefont {Segal}, \citenamefont {Plotnik}, \citenamefont {Shalaev},\ and\ \citenamefont {Segev}}]{Konforty2025}%
  \BibitemOpen
  \bibfield  {author} {\bibinfo {author} {\bibfnamefont {N.}~\bibnamefont {Konforty}}, \bibinfo {author} {\bibfnamefont {M.-I.}\ \bibnamefont {Cohen}}, \bibinfo {author} {\bibfnamefont {O.}~\bibnamefont {Segal}}, \bibinfo {author} {\bibfnamefont {Y.}~\bibnamefont {Plotnik}}, \bibinfo {author} {\bibfnamefont {V.~M.}\ \bibnamefont {Shalaev}},\ and\ \bibinfo {author} {\bibfnamefont {M.}~\bibnamefont {Segev}},\ }\bibfield  {title} {\bibinfo {title} {Second harmonic generation and nonlinear frequency conversion in photonic time-crystals},\ }\href {https://doi.org/10.1038/s41377-025-01788-z} {\bibfield  {journal} {\bibinfo  {journal} {Light Sci. Appl.}\ }\textbf {\bibinfo {volume} {14}},\ \bibinfo {pages} {152} (\bibinfo {year} {2025})}\BibitemShut {NoStop}%
\bibitem [{\citenamefont {Dong}\ \emph {et~al.}(2025{\natexlab{a}})\citenamefont {Dong}, \citenamefont {Chen},\ and\ \citenamefont {Yuan}}]{DongZH2025}%
  \BibitemOpen
  \bibfield  {author} {\bibinfo {author} {\bibfnamefont {Z.}~\bibnamefont {Dong}}, \bibinfo {author} {\bibfnamefont {X.}~\bibnamefont {Chen}},\ and\ \bibinfo {author} {\bibfnamefont {L.}~\bibnamefont {Yuan}},\ }\bibfield  {title} {\bibinfo {title} {Extremely narrow band in moir\'e photonic time crystal},\ }\href {https://doi.org/10.1103/4lqd-z567} {\bibfield  {journal} {\bibinfo  {journal} {Phys. Rev. Lett.}\ }\textbf {\bibinfo {volume} {135}},\ \bibinfo {pages} {033803} (\bibinfo {year} {2025}{\natexlab{a}})}\BibitemShut {NoStop}%
\bibitem [{\citenamefont {Sharabi}\ \emph {et~al.}(2021)\citenamefont {Sharabi}, \citenamefont {Lustig},\ and\ \citenamefont {Segev}}]{Sharabi2021}%
  \BibitemOpen
  \bibfield  {author} {\bibinfo {author} {\bibfnamefont {Y.}~\bibnamefont {Sharabi}}, \bibinfo {author} {\bibfnamefont {E.}~\bibnamefont {Lustig}},\ and\ \bibinfo {author} {\bibfnamefont {M.}~\bibnamefont {Segev}},\ }\bibfield  {title} {\bibinfo {title} {Disordered photonic time crystals},\ }\href {https://doi.org/10.1103/PhysRevLett.126.163902} {\bibfield  {journal} {\bibinfo  {journal} {Phys. Rev. Lett.}\ }\textbf {\bibinfo {volume} {126}},\ \bibinfo {pages} {163902} (\bibinfo {year} {2021})}\BibitemShut {NoStop}%
\bibitem [{\citenamefont {Kim}\ \emph {et~al.}(2023)\citenamefont {Kim}, \citenamefont {Lee}, \citenamefont {Yu},\ and\ \citenamefont {Park}}]{KimJM2023}%
  \BibitemOpen
  \bibfield  {author} {\bibinfo {author} {\bibfnamefont {J.}~\bibnamefont {Kim}}, \bibinfo {author} {\bibfnamefont {D.}~\bibnamefont {Lee}}, \bibinfo {author} {\bibfnamefont {S.}~\bibnamefont {Yu}},\ and\ \bibinfo {author} {\bibfnamefont {N.}~\bibnamefont {Park}},\ }\bibfield  {title} {\bibinfo {title} {Unidirectional scattering with spatial homogeneity using correlated photonic time disorder},\ }\href@noop {} {\bibfield  {journal} {\bibinfo  {journal} {Nat. Phys.}\ }\textbf {\bibinfo {volume} {19}},\ \bibinfo {pages} {726} (\bibinfo {year} {2023})}\BibitemShut {NoStop}%
\bibitem [{\citenamefont {Galiffi}\ \emph {et~al.}(2020)\citenamefont {Galiffi}, \citenamefont {Wang}, \citenamefont {Lim}, \citenamefont {Pendry}, \citenamefont {Al\`u},\ and\ \citenamefont {Huidobro}}]{Wood2020}%
  \BibitemOpen
  \bibfield  {author} {\bibinfo {author} {\bibfnamefont {E.}~\bibnamefont {Galiffi}}, \bibinfo {author} {\bibfnamefont {Y.-T.}\ \bibnamefont {Wang}}, \bibinfo {author} {\bibfnamefont {Z.}~\bibnamefont {Lim}}, \bibinfo {author} {\bibfnamefont {J.~B.}\ \bibnamefont {Pendry}}, \bibinfo {author} {\bibfnamefont {A.}~\bibnamefont {Al\`u}},\ and\ \bibinfo {author} {\bibfnamefont {P.~A.}\ \bibnamefont {Huidobro}},\ }\bibfield  {title} {\bibinfo {title} {Wood anomalies and surface-wave excitation with a time grating},\ }\href {https://doi.org/10.1103/PhysRevLett.125.127403} {\bibfield  {journal} {\bibinfo  {journal} {Phys. Rev. Lett.}\ }\textbf {\bibinfo {volume} {125}},\ \bibinfo {pages} {127403} (\bibinfo {year} {2020})}\BibitemShut {NoStop}%
\bibitem [{\citenamefont {Sharabi}\ \emph {et~al.}(2022)\citenamefont {Sharabi}, \citenamefont {Dikopoltsev}, \citenamefont {Lustig}, \citenamefont {Lumer},\ and\ \citenamefont {Segev}}]{Sharabi2022Optica}%
  \BibitemOpen
  \bibfield  {author} {\bibinfo {author} {\bibfnamefont {Y.}~\bibnamefont {Sharabi}}, \bibinfo {author} {\bibfnamefont {A.}~\bibnamefont {Dikopoltsev}}, \bibinfo {author} {\bibfnamefont {E.}~\bibnamefont {Lustig}}, \bibinfo {author} {\bibfnamefont {Y.}~\bibnamefont {Lumer}},\ and\ \bibinfo {author} {\bibfnamefont {M.}~\bibnamefont {Segev}},\ }\bibfield  {title} {\bibinfo {title} {Spatiotemporal photonic crystals},\ }\href {https://doi.org/10.1364/OPTICA.455672} {\bibfield  {journal} {\bibinfo  {journal} {Optica}\ }\textbf {\bibinfo {volume} {9}},\ \bibinfo {pages} {585} (\bibinfo {year} {2022})}\BibitemShut {NoStop}%
\bibitem [{\citenamefont {Wang}\ \emph {et~al.}(2023)\citenamefont {Wang}, \citenamefont {Mirmoosa}, \citenamefont {Asadchy}, \citenamefont {Rockstuhl}, \citenamefont {Fan},\ and\ \citenamefont {Tretyakov}}]{Wang2023}%
  \BibitemOpen
  \bibfield  {author} {\bibinfo {author} {\bibfnamefont {X.}~\bibnamefont {Wang}}, \bibinfo {author} {\bibfnamefont {M.~S.}\ \bibnamefont {Mirmoosa}}, \bibinfo {author} {\bibfnamefont {V.~S.}\ \bibnamefont {Asadchy}}, \bibinfo {author} {\bibfnamefont {C.}~\bibnamefont {Rockstuhl}}, \bibinfo {author} {\bibfnamefont {S.}~\bibnamefont {Fan}},\ and\ \bibinfo {author} {\bibfnamefont {S.~A.}\ \bibnamefont {Tretyakov}},\ }\bibfield  {title} {\bibinfo {title} {{Metasurface-based realization of photonic time crystals}},\ }\href {https://doi.org/10.1126/sciadv.adg7541} {\bibfield  {journal} {\bibinfo  {journal} {Sci. Adv.}\ }\textbf {\bibinfo {volume} {9}},\ \bibinfo {pages} {eadg7541} (\bibinfo {year} {2023})}\BibitemShut {NoStop}%
\bibitem [{\citenamefont {Moussa}\ \emph {et~al.}(2023)\citenamefont {Moussa}, \citenamefont {Xu}, \citenamefont {Yin}, \citenamefont {Galiffi}, \citenamefont {Ra'di},\ and\ \citenamefont {Al\'u}}]{Moussa2023}%
  \BibitemOpen
  \bibfield  {author} {\bibinfo {author} {\bibfnamefont {H.}~\bibnamefont {Moussa}}, \bibinfo {author} {\bibfnamefont {G.}~\bibnamefont {Xu}}, \bibinfo {author} {\bibfnamefont {S.}~\bibnamefont {Yin}}, \bibinfo {author} {\bibfnamefont {E.}~\bibnamefont {Galiffi}}, \bibinfo {author} {\bibfnamefont {Y.}~\bibnamefont {Ra'di}},\ and\ \bibinfo {author} {\bibfnamefont {A.}~\bibnamefont {Al\'u}},\ }\bibfield  {title} {\bibinfo {title} {{Observation of temporal reflection and broadband frequency translation at photonic time interfaces}},\ }\href {http://dx.doi.org/10.1038/s41567-023-01975-y} {\bibfield  {journal} {\bibinfo  {journal} {Nat. Phys.}\ }\textbf {\bibinfo {volume} {19}},\ \bibinfo {pages} {863} (\bibinfo {year} {2023})}\BibitemShut {NoStop}%
\bibitem [{\citenamefont {Jones}\ \emph {et~al.}(2024)\citenamefont {Jones}, \citenamefont {Kildishev}, \citenamefont {Segev},\ and\ \citenamefont {Peroulis}}]{Jones2024}%
  \BibitemOpen
  \bibfield  {author} {\bibinfo {author} {\bibfnamefont {T.}~\bibnamefont {Jones}}, \bibinfo {author} {\bibfnamefont {A.}~\bibnamefont {Kildishev}}, \bibinfo {author} {\bibfnamefont {M.}~\bibnamefont {Segev}},\ and\ \bibinfo {author} {\bibfnamefont {D.}~\bibnamefont {Peroulis}},\ }\bibfield  {title} {\bibinfo {title} {{Time-reflection of microwaves by a fast optically-controlled time-boundary}},\ }\href {https://doi.org/10.1038/s41467-024-51171-6} {\bibfield  {journal} {\bibinfo  {journal} {Nat. Commun.}\ }\textbf {\bibinfo {volume} {15}},\ \bibinfo {pages} {6786} (\bibinfo {year} {2024})}\BibitemShut {NoStop}%
\bibitem [{\citenamefont {Alam}\ \emph {et~al.}(2016)\citenamefont {Alam}, \citenamefont {Leon},\ and\ \citenamefont {Boyd}}]{Alam2016}%
  \BibitemOpen
  \bibfield  {author} {\bibinfo {author} {\bibfnamefont {M.~Z.}\ \bibnamefont {Alam}}, \bibinfo {author} {\bibfnamefont {I.~D.}\ \bibnamefont {Leon}},\ and\ \bibinfo {author} {\bibfnamefont {R.~W.}\ \bibnamefont {Boyd}},\ }\bibfield  {title} {\bibinfo {title} {{Large optical nonlinearity of indium tin oxide in its epsilon-near-zero region}},\ }\href {https://doi.org/10.1126/science.aae0330} {\bibfield  {journal} {\bibinfo  {journal} {Science}\ }\textbf {\bibinfo {volume} {352}},\ \bibinfo {pages} {795} (\bibinfo {year} {2016})}\BibitemShut {NoStop}%
\bibitem [{\citenamefont {Tirole}\ \emph {et~al.}(2023)\citenamefont {Tirole}, \citenamefont {Vezzoli}, \citenamefont {Galiffi}, \citenamefont {Robertson}, \citenamefont {Maurice}, \citenamefont {Tilmann}, \citenamefont {Maier}, \citenamefont {Pendry},\ and\ \citenamefont {Sapienza}}]{Tirole2023}%
  \BibitemOpen
  \bibfield  {author} {\bibinfo {author} {\bibfnamefont {R.}~\bibnamefont {Tirole}}, \bibinfo {author} {\bibfnamefont {S.}~\bibnamefont {Vezzoli}}, \bibinfo {author} {\bibfnamefont {E.}~\bibnamefont {Galiffi}}, \bibinfo {author} {\bibfnamefont {I.}~\bibnamefont {Robertson}}, \bibinfo {author} {\bibfnamefont {D.}~\bibnamefont {Maurice}}, \bibinfo {author} {\bibfnamefont {B.}~\bibnamefont {Tilmann}}, \bibinfo {author} {\bibfnamefont {S.~A.}\ \bibnamefont {Maier}}, \bibinfo {author} {\bibfnamefont {J.~B.}\ \bibnamefont {Pendry}},\ and\ \bibinfo {author} {\bibfnamefont {R.}~\bibnamefont {Sapienza}},\ }\bibfield  {title} {\bibinfo {title} {Double-slit time diffraction at optical frequencies},\ }\href {https://doi.org/10.1038/s41567-023-01993-w} {\bibfield  {journal} {\bibinfo  {journal} {Nat. Phys.}\ }\textbf {\bibinfo {volume} {19}},\ \bibinfo {pages} {999} (\bibinfo {year} {2023})}\BibitemShut {NoStop}%
\bibitem [{\citenamefont {Zhou}\ \emph {et~al.}(2020)\citenamefont {Zhou}, \citenamefont {Alam}, \citenamefont {Karimi}, \citenamefont {Upham}, \citenamefont {Reshef}, \citenamefont {Liu}, \citenamefont {Willner},\ and\ \citenamefont {Boyd}}]{Zhou2020}%
  \BibitemOpen
  \bibfield  {author} {\bibinfo {author} {\bibfnamefont {Y.}~\bibnamefont {Zhou}}, \bibinfo {author} {\bibfnamefont {M.~Z.}\ \bibnamefont {Alam}}, \bibinfo {author} {\bibfnamefont {M.}~\bibnamefont {Karimi}}, \bibinfo {author} {\bibfnamefont {J.}~\bibnamefont {Upham}}, \bibinfo {author} {\bibfnamefont {O.}~\bibnamefont {Reshef}}, \bibinfo {author} {\bibfnamefont {C.}~\bibnamefont {Liu}}, \bibinfo {author} {\bibfnamefont {A.~E.}\ \bibnamefont {Willner}},\ and\ \bibinfo {author} {\bibfnamefont {R.~W.}\ \bibnamefont {Boyd}},\ }\bibfield  {title} {\bibinfo {title} {{Broadband frequency translation through time refraction in an epsilon-near-zero material.}},\ }\href {https://doi.org/10.1038/s41467-020-15682-2} {\bibfield  {journal} {\bibinfo  {journal} {Nat. Commun.}\ }\textbf {\bibinfo {volume} {11}},\ \bibinfo {pages} {2180} (\bibinfo {year} {2020})}\BibitemShut {NoStop}%
\bibitem [{\citenamefont {Bohn}\ \emph {et~al.}(2021)\citenamefont {Bohn}, \citenamefont {Luk}, \citenamefont {Horsley},\ and\ \citenamefont {Hendry}}]{Bohn2021}%
  \BibitemOpen
  \bibfield  {author} {\bibinfo {author} {\bibfnamefont {J.}~\bibnamefont {Bohn}}, \bibinfo {author} {\bibfnamefont {T.~S.}\ \bibnamefont {Luk}}, \bibinfo {author} {\bibfnamefont {S.}~\bibnamefont {Horsley}},\ and\ \bibinfo {author} {\bibfnamefont {E.}~\bibnamefont {Hendry}},\ }\bibfield  {title} {\bibinfo {title} {{Spatiotemporal refraction of light in an epsilon-near-zero indium tin oxide layer: frequency shifting effects arising from interfaces}},\ }\href {https://doi.org/10.1364/optica.436324} {\bibfield  {journal} {\bibinfo  {journal} {Optica}\ }\textbf {\bibinfo {volume} {8}},\ \bibinfo {pages} {1532} (\bibinfo {year} {2021})}\BibitemShut {NoStop}%
\bibitem [{\citenamefont {Lustig}\ \emph {et~al.}(2023)\citenamefont {Lustig}, \citenamefont {Segal}, \citenamefont {Saha}, \citenamefont {Bordo}, \citenamefont {Chowdhury}, \citenamefont {Sharabi}, \citenamefont {Fleischer}, \citenamefont {Boltasseva}, \citenamefont {Cohen}, \citenamefont {Shalaev},\ and\ \citenamefont {Segev}}]{Eran2023}%
  \BibitemOpen
  \bibfield  {author} {\bibinfo {author} {\bibfnamefont {E.}~\bibnamefont {Lustig}}, \bibinfo {author} {\bibfnamefont {O.}~\bibnamefont {Segal}}, \bibinfo {author} {\bibfnamefont {S.}~\bibnamefont {Saha}}, \bibinfo {author} {\bibfnamefont {E.}~\bibnamefont {Bordo}}, \bibinfo {author} {\bibfnamefont {S.~N.}\ \bibnamefont {Chowdhury}}, \bibinfo {author} {\bibfnamefont {Y.}~\bibnamefont {Sharabi}}, \bibinfo {author} {\bibfnamefont {A.}~\bibnamefont {Fleischer}}, \bibinfo {author} {\bibfnamefont {A.}~\bibnamefont {Boltasseva}}, \bibinfo {author} {\bibfnamefont {O.}~\bibnamefont {Cohen}}, \bibinfo {author} {\bibfnamefont {V.~M.}\ \bibnamefont {Shalaev}},\ and\ \bibinfo {author} {\bibfnamefont {M.}~\bibnamefont {Segev}},\ }\bibfield  {title} {\bibinfo {title} {{Time-refraction optics with single cycle modulation}},\ }\href {https://doi.org/10.1515/nanoph-2023-0126} {\bibfield  {journal} {\bibinfo  {journal} {Nanophotonics}\ }\textbf {\bibinfo {volume} {12}},\ \bibinfo {pages} {2221} (\bibinfo {year}
  {2023})}\BibitemShut {NoStop}%
\bibitem [{\citenamefont {Lu}\ \emph {et~al.}(2025)\citenamefont {Lu}, \citenamefont {Zhang}, \citenamefont {Qiu}, \citenamefont {Niu}, \citenamefont {Chen}, \citenamefont {Xu}, \citenamefont {Zhang}, \citenamefont {Zhang},\ and\ \citenamefont {Han}}]{Lu2025}%
  \BibitemOpen
  \bibfield  {author} {\bibinfo {author} {\bibfnamefont {Y.}~\bibnamefont {Lu}}, \bibinfo {author} {\bibfnamefont {X.}~\bibnamefont {Zhang}}, \bibinfo {author} {\bibfnamefont {H.}~\bibnamefont {Qiu}}, \bibinfo {author} {\bibfnamefont {L.}~\bibnamefont {Niu}}, \bibinfo {author} {\bibfnamefont {X.}~\bibnamefont {Chen}}, \bibinfo {author} {\bibfnamefont {Q.}~\bibnamefont {Xu}}, \bibinfo {author} {\bibfnamefont {W.}~\bibnamefont {Zhang}}, \bibinfo {author} {\bibfnamefont {S.}~\bibnamefont {Zhang}},\ and\ \bibinfo {author} {\bibfnamefont {J.}~\bibnamefont {Han}},\ }\bibfield  {title} {\bibinfo {title} {{Ultrafast Temporal Modulation of Terahertz Generation at an Optically Pumped ITO Interface}},\ }\href {10.1364/OPTICA.555318} {\bibfield  {journal} {\bibinfo  {journal} {Optica}\ }\textbf {\bibinfo {volume} {12}},\ \bibinfo {pages} {1035} (\bibinfo {year} {2025})}\BibitemShut {NoStop}%
\bibitem [{\citenamefont {Zheng}\ \emph {et~al.}(2017)\citenamefont {Zheng}, \citenamefont {Huang}, \citenamefont {Guo}, \citenamefont {Meng}, \citenamefont {Lu}, \citenamefont {Wang}, \citenamefont {Zhao}, \citenamefont {Meng}, \citenamefont {Zhang}, \citenamefont {Yuan} \emph {et~al.}}]{zheng2017filament}%
  \BibitemOpen
  \bibfield  {author} {\bibinfo {author} {\bibfnamefont {Z.}~\bibnamefont {Zheng}}, \bibinfo {author} {\bibfnamefont {Y.}~\bibnamefont {Huang}}, \bibinfo {author} {\bibfnamefont {Q.}~\bibnamefont {Guo}}, \bibinfo {author} {\bibfnamefont {C.}~\bibnamefont {Meng}}, \bibinfo {author} {\bibfnamefont {Z.}~\bibnamefont {Lu}}, \bibinfo {author} {\bibfnamefont {X.}~\bibnamefont {Wang}}, \bibinfo {author} {\bibfnamefont {J.}~\bibnamefont {Zhao}}, \bibinfo {author} {\bibfnamefont {C.}~\bibnamefont {Meng}}, \bibinfo {author} {\bibfnamefont {D.}~\bibnamefont {Zhang}}, \bibinfo {author} {\bibfnamefont {J.}~\bibnamefont {Yuan}}, \emph {et~al.},\ }\bibfield  {title} {\bibinfo {title} {Filament characterization via resonance absorption of terahertz wave},\ }\href@noop {} {\bibfield  {journal} {\bibinfo  {journal} {Phys. Plasmas}\ }\textbf {\bibinfo {volume} {24}},\ \bibinfo {pages} {103303} (\bibinfo {year} {2017})}\BibitemShut {NoStop}%
\bibitem [{\citenamefont {Qu}\ \emph {et~al.}(2024)\citenamefont {Qu}, \citenamefont {Huang}, \citenamefont {Zhou}, \citenamefont {Gao}, \citenamefont {Lou}, \citenamefont {Feng}, \citenamefont {Zhao}, \citenamefont {Chang}, \citenamefont {Shkurinov},\ and\ \citenamefont {Verboncoeur}}]{Qu2024}%
  \BibitemOpen
  \bibfield  {author} {\bibinfo {author} {\bibfnamefont {X.}~\bibnamefont {Qu}}, \bibinfo {author} {\bibfnamefont {Y.}~\bibnamefont {Huang}}, \bibinfo {author} {\bibfnamefont {B.}~\bibnamefont {Zhou}}, \bibinfo {author} {\bibfnamefont {M.}~\bibnamefont {Gao}}, \bibinfo {author} {\bibfnamefont {J.}~\bibnamefont {Lou}}, \bibinfo {author} {\bibfnamefont {Y.}~\bibnamefont {Feng}}, \bibinfo {author} {\bibfnamefont {Z.}~\bibnamefont {Zhao}}, \bibinfo {author} {\bibfnamefont {C.}~\bibnamefont {Chang}}, \bibinfo {author} {\bibfnamefont {A.~P.}\ \bibnamefont {Shkurinov}},\ and\ \bibinfo {author} {\bibfnamefont {J.}~\bibnamefont {Verboncoeur}},\ }\bibfield  {title} {\bibinfo {title} {Ultrafast plasma-based terahertz modulator},\ }\href@noop {} {\bibfield  {journal} {\bibinfo  {journal} {Optica}\ }\textbf {\bibinfo {volume} {11}},\ \bibinfo {pages} {1478} (\bibinfo {year} {2024})}\BibitemShut {NoStop}%
\bibitem [{\citenamefont {Wang}\ \emph {et~al.}(2022)\citenamefont {Wang}, \citenamefont {Ye}, \citenamefont {Sun}, \citenamefont {Han}, \citenamefont {Hou},\ and\ \citenamefont {Zhang}}]{WangXK2022}%
  \BibitemOpen
  \bibfield  {author} {\bibinfo {author} {\bibfnamefont {X.-K.}\ \bibnamefont {Wang}}, \bibinfo {author} {\bibfnamefont {J.-S.}\ \bibnamefont {Ye}}, \bibinfo {author} {\bibfnamefont {W.-F.}\ \bibnamefont {Sun}}, \bibinfo {author} {\bibfnamefont {P.}~\bibnamefont {Han}}, \bibinfo {author} {\bibfnamefont {L.}~\bibnamefont {Hou}},\ and\ \bibinfo {author} {\bibfnamefont {Y.}~\bibnamefont {Zhang}},\ }\bibfield  {title} {\bibinfo {title} {Terahertz near-field microscopy based on an air-plasma dynamic aperture},\ }\href@noop {} {\bibfield  {journal} {\bibinfo  {journal} {Light Sci. Appl.}\ }\textbf {\bibinfo {volume} {11}},\ \bibinfo {pages} {129} (\bibinfo {year} {2022})}\BibitemShut {NoStop}%
\bibitem [{\citenamefont {Zhao}\ \emph {et~al.}(2023)\citenamefont {Zhao}, \citenamefont {Zhu}, \citenamefont {Han}, \citenamefont {Wang}, \citenamefont {Lao}, \citenamefont {Li}, \citenamefont {Peng},\ and\ \citenamefont {Zhu}}]{ZhaoJY2023}%
  \BibitemOpen
  \bibfield  {author} {\bibinfo {author} {\bibfnamefont {J.}~\bibnamefont {Zhao}}, \bibinfo {author} {\bibfnamefont {F.}~\bibnamefont {Zhu}}, \bibinfo {author} {\bibfnamefont {Y.}~\bibnamefont {Han}}, \bibinfo {author} {\bibfnamefont {Q.}~\bibnamefont {Wang}}, \bibinfo {author} {\bibfnamefont {L.}~\bibnamefont {Lao}}, \bibinfo {author} {\bibfnamefont {X.}~\bibnamefont {Li}}, \bibinfo {author} {\bibfnamefont {Y.}~\bibnamefont {Peng}},\ and\ \bibinfo {author} {\bibfnamefont {Y.}~\bibnamefont {Zhu}},\ }\bibfield  {title} {\bibinfo {title} {{Light-guiding-light-based temporal integration of broadband terahertz pulses in air}},\ }\href {https://doi.org/10.1063/5.0158107} {\bibfield  {journal} {\bibinfo  {journal} {APL Photonics}\ }\textbf {\bibinfo {volume} {8}},\ \bibinfo {pages} {106107} (\bibinfo {year} {2023})}\BibitemShut {NoStop}%
\bibitem [{\citenamefont {Huang}\ \emph {et~al.}(2021)\citenamefont {Huang}, \citenamefont {Xiang}, \citenamefont {Xu}, \citenamefont {Zhao}, \citenamefont {Liu}, \citenamefont {Wang}, \citenamefont {Zhang}, \citenamefont {L\"u}, \citenamefont {Zhang}, \citenamefont {Chang}, \citenamefont {Yuan},\ and\ \citenamefont {Zhao}}]{HuangYD2021PRA}%
  \BibitemOpen
  \bibfield  {author} {\bibinfo {author} {\bibfnamefont {Y.}~\bibnamefont {Huang}}, \bibinfo {author} {\bibfnamefont {Z.}~\bibnamefont {Xiang}}, \bibinfo {author} {\bibfnamefont {X.}~\bibnamefont {Xu}}, \bibinfo {author} {\bibfnamefont {J.}~\bibnamefont {Zhao}}, \bibinfo {author} {\bibfnamefont {J.}~\bibnamefont {Liu}}, \bibinfo {author} {\bibfnamefont {R.}~\bibnamefont {Wang}}, \bibinfo {author} {\bibfnamefont {Z.}~\bibnamefont {Zhang}}, \bibinfo {author} {\bibfnamefont {Z.}~\bibnamefont {L\"u}}, \bibinfo {author} {\bibfnamefont {D.}~\bibnamefont {Zhang}}, \bibinfo {author} {\bibfnamefont {C.}~\bibnamefont {Chang}}, \bibinfo {author} {\bibfnamefont {J.}~\bibnamefont {Yuan}},\ and\ \bibinfo {author} {\bibfnamefont {Z.}~\bibnamefont {Zhao}},\ }\bibfield  {title} {\bibinfo {title} {Localized-plasma-assisted rotational transitions in the terahertz region},\ }\href@noop {} {\bibfield  {journal} {\bibinfo  {journal} {Phys. Rev. A}\ }\textbf {\bibinfo {volume} {103}},\ \bibinfo {pages} {033109} (\bibinfo {year}
  {2021})}\BibitemShut {NoStop}%
\bibitem [{\citenamefont {Tong}\ \emph {et~al.}(2002)\citenamefont {Tong}, \citenamefont {Zhao},\ and\ \citenamefont {Lin}}]{TongXM2002}%
  \BibitemOpen
  \bibfield  {author} {\bibinfo {author} {\bibfnamefont {X.~M.}\ \bibnamefont {Tong}}, \bibinfo {author} {\bibfnamefont {Z.~X.}\ \bibnamefont {Zhao}},\ and\ \bibinfo {author} {\bibfnamefont {C.~D.}\ \bibnamefont {Lin}},\ }\bibfield  {title} {\bibinfo {title} {{Theory of molecular tunneling ionization}},\ }\href {https://doi.org/10.1103/PhysRevA.66.033402} {\bibfield  {journal} {\bibinfo  {journal} {Phys. Rev. A}\ }\textbf {\bibinfo {volume} {66}},\ \bibinfo {pages} {033402} (\bibinfo {year} {2002})}\BibitemShut {NoStop}%
\bibitem [{\citenamefont {{Smirnov}}\ and\ \citenamefont {{Chibisov}}(1966)}]{Smirnov1966}%
  \BibitemOpen
  \bibfield  {author} {\bibinfo {author} {\bibfnamefont {B.~M.}\ \bibnamefont {{Smirnov}}}\ and\ \bibinfo {author} {\bibfnamefont {M.~I.}\ \bibnamefont {{Chibisov}}},\ }\bibfield  {title} {\bibinfo {title} {{The Breaking Up of Atomic Particles by an Electric Field and by Electron Collisions}},\ }\href@noop {} {\bibfield  {journal} {\bibinfo  {journal} {Sov. Phys. JETP}\ }\textbf {\bibinfo {volume} {22}},\ \bibinfo {pages} {585} (\bibinfo {year} {1966})}\BibitemShut {NoStop}%
\bibitem [{\citenamefont {Ammosov}\ \emph {et~al.}(1986)\citenamefont {Ammosov}, \citenamefont {Delone},\ and\ \citenamefont {Krainov}}]{Ammosov1986}%
  \BibitemOpen
  \bibfield  {author} {\bibinfo {author} {\bibfnamefont {M.-V.}\ \bibnamefont {Ammosov}}, \bibinfo {author} {\bibfnamefont {N.-B.}\ \bibnamefont {Delone}},\ and\ \bibinfo {author} {\bibfnamefont {V.-P.}\ \bibnamefont {Krainov}},\ }\bibfield  {title} {\bibinfo {title} {Tunnel ionization of complex atoms and atomic ions in electromagnetic field},\ }\href@noop {} {\bibfield  {journal} {\bibinfo  {journal} {Sov. Phys. JETP}\ }\textbf {\bibinfo {volume} {64}},\ \bibinfo {pages} {1191} (\bibinfo {year} {1986})}\BibitemShut {NoStop}%
\bibitem [{\citenamefont {Mics}\ \emph {et~al.}(2005)\citenamefont {Mics}, \citenamefont {Kadlec}, \citenamefont {Ku\v{z}el}, \citenamefont {Jungwirth}, \citenamefont {Bradforth},\ and\ \citenamefont {Apkarian}}]{mics2005nonresonant}%
  \BibitemOpen
  \bibfield  {author} {\bibinfo {author} {\bibfnamefont {Z.}~\bibnamefont {Mics}}, \bibinfo {author} {\bibfnamefont {F.}~\bibnamefont {Kadlec}}, \bibinfo {author} {\bibfnamefont {P.}~\bibnamefont {Ku\v{z}el}}, \bibinfo {author} {\bibfnamefont {P.}~\bibnamefont {Jungwirth}}, \bibinfo {author} {\bibfnamefont {S.~E.}\ \bibnamefont {Bradforth}},\ and\ \bibinfo {author} {\bibfnamefont {V.~A.}\ \bibnamefont {Apkarian}},\ }\bibfield  {title} {\bibinfo {title} {Nonresonant ionization of oxygen molecules by femtosecond pulses: Plasma dynamics studied by time-resolved terahertz spectroscopy},\ }\href@noop {} {\bibfield  {journal} {\bibinfo  {journal} {J. Chem. Phys.}\ }\textbf {\bibinfo {volume} {123}},\ \bibinfo {pages} {104310} (\bibinfo {year} {2005})}\BibitemShut {NoStop}%
\bibitem [{\citenamefont {Sol\'{i}s}\ \emph {et~al.}(2021)\citenamefont {Sol\'{i}s}, \citenamefont {Kastner},\ and\ \citenamefont {Engheta}}]{Solis2021}%
  \BibitemOpen
  \bibfield  {author} {\bibinfo {author} {\bibfnamefont {D.~M.}\ \bibnamefont {Sol\'{i}s}}, \bibinfo {author} {\bibfnamefont {R.}~\bibnamefont {Kastner}},\ and\ \bibinfo {author} {\bibfnamefont {N.}~\bibnamefont {Engheta}},\ }\bibfield  {title} {\bibinfo {title} {Time-varying materials in the presence of dispersion: plane-wave propagation in a lorentzian medium with temporal discontinuity},\ }\href {https://doi.org/10.1364/PRJ.427368} {\bibfield  {journal} {\bibinfo  {journal} {Photonics Res.}\ }\textbf {\bibinfo {volume} {9}},\ \bibinfo {pages} {1842} (\bibinfo {year} {2021})}\BibitemShut {NoStop}%
\bibitem [{\citenamefont {Galiffi}\ \emph {et~al.}(2022)\citenamefont {Galiffi}, \citenamefont {Tirole}, \citenamefont {Yin}, \citenamefont {Li}, \citenamefont {Vezzoli}, \citenamefont {Huidobro}, \citenamefont {Silveirinha}, \citenamefont {Sapienza}, \citenamefont {Al{\`u}},\ and\ \citenamefont {Pendry}}]{Galiffi2022}%
  \BibitemOpen
  \bibfield  {author} {\bibinfo {author} {\bibfnamefont {E.}~\bibnamefont {Galiffi}}, \bibinfo {author} {\bibfnamefont {R.}~\bibnamefont {Tirole}}, \bibinfo {author} {\bibfnamefont {S.}~\bibnamefont {Yin}}, \bibinfo {author} {\bibfnamefont {H.}~\bibnamefont {Li}}, \bibinfo {author} {\bibfnamefont {S.}~\bibnamefont {Vezzoli}}, \bibinfo {author} {\bibfnamefont {P.~A.}\ \bibnamefont {Huidobro}}, \bibinfo {author} {\bibfnamefont {M.~G.}\ \bibnamefont {Silveirinha}}, \bibinfo {author} {\bibfnamefont {R.}~\bibnamefont {Sapienza}}, \bibinfo {author} {\bibfnamefont {A.}~\bibnamefont {Al{\`u}}},\ and\ \bibinfo {author} {\bibfnamefont {J.~B.}\ \bibnamefont {Pendry}},\ }\bibfield  {title} {\bibinfo {title} {{Photonics of time-varying media}},\ }\href {https://doi.org/10.1117/1.AP.4.1.014002} {\bibfield  {journal} {\bibinfo  {journal} {Adv. Photonics}\ }\textbf {\bibinfo {volume} {4}},\ \bibinfo {pages} {014002} (\bibinfo {year} {2022})}\BibitemShut {NoStop}%
\bibitem [{\citenamefont {Schr{\"o}del}\ \emph {et~al.}(2023)\citenamefont {Schr{\"o}del}, \citenamefont {Hartmann}, \citenamefont {Zheng}, \citenamefont {Lang}, \citenamefont {Steudel}, \citenamefont {Rutsch}, \citenamefont {Salman}, \citenamefont {Kellert}, \citenamefont {Pergament}, \citenamefont {Hahn-Jose}, \citenamefont {Suppelt}, \citenamefont {D{\"o}rsam}, \citenamefont {Harth}, \citenamefont {Leemans}, \citenamefont {K{\"a}rtner}, \citenamefont {Hartl}, \citenamefont {Kupnik},\ and\ \citenamefont {Heyl}}]{Schrodel2023AcoustoopticMO}%
  \BibitemOpen
  \bibfield  {author} {\bibinfo {author} {\bibfnamefont {Y.}~\bibnamefont {Schr{\"o}del}}, \bibinfo {author} {\bibfnamefont {C.}~\bibnamefont {Hartmann}}, \bibinfo {author} {\bibfnamefont {J.}~\bibnamefont {Zheng}}, \bibinfo {author} {\bibfnamefont {T.}~\bibnamefont {Lang}}, \bibinfo {author} {\bibfnamefont {M.}~\bibnamefont {Steudel}}, \bibinfo {author} {\bibfnamefont {M.}~\bibnamefont {Rutsch}}, \bibinfo {author} {\bibfnamefont {S.~H.}\ \bibnamefont {Salman}}, \bibinfo {author} {\bibfnamefont {M.}~\bibnamefont {Kellert}}, \bibinfo {author} {\bibfnamefont {M.}~\bibnamefont {Pergament}}, \bibinfo {author} {\bibfnamefont {T.}~\bibnamefont {Hahn-Jose}}, \bibinfo {author} {\bibfnamefont {S.}~\bibnamefont {Suppelt}}, \bibinfo {author} {\bibfnamefont {J.~H.}\ \bibnamefont {D{\"o}rsam}}, \bibinfo {author} {\bibfnamefont {A.}~\bibnamefont {Harth}}, \bibinfo {author} {\bibfnamefont {W.~P.}\ \bibnamefont {Leemans}}, \bibinfo {author} {\bibfnamefont {F.~X.}\ \bibnamefont {K{\"a}rtner}}, \bibinfo {author} {\bibfnamefont
  {I.}~\bibnamefont {Hartl}}, \bibinfo {author} {\bibfnamefont {M.}~\bibnamefont {Kupnik}},\ and\ \bibinfo {author} {\bibfnamefont {C.~M.}\ \bibnamefont {Heyl}},\ }\bibfield  {title} {\bibinfo {title} {Acousto-optic modulation of gigawatt-scale laser pulses in ambient air},\ }\href {https://api.semanticscholar.org/CorpusID:263657311} {\bibfield  {journal} {\bibinfo  {journal} {Nat. Photonics}\ }\textbf {\bibinfo {volume} {18}},\ \bibinfo {pages} {54} (\bibinfo {year} {2023})}\BibitemShut {NoStop}%
\bibitem [{\citenamefont {Zhang}\ \emph {et~al.}(2025)\citenamefont {Zhang}, \citenamefont {Weng}, \citenamefont {Ai}, \citenamefont {Qiao}, \citenamefont {Xue},\ and\ \citenamefont {Sheng}}]{ZhangXB2025}%
  \BibitemOpen
  \bibfield  {author} {\bibinfo {author} {\bibfnamefont {X.-B.}\ \bibnamefont {Zhang}}, \bibinfo {author} {\bibfnamefont {S.-M.}\ \bibnamefont {Weng}}, \bibinfo {author} {\bibfnamefont {H.}~\bibnamefont {Ai}}, \bibinfo {author} {\bibfnamefont {X.}~\bibnamefont {Qiao}}, \bibinfo {author} {\bibfnamefont {J.-K.}\ \bibnamefont {Xue}},\ and\ \bibinfo {author} {\bibfnamefont {Z.-M.}\ \bibnamefont {Sheng}},\ }\bibfield  {title} {\bibinfo {title} {Photonic rabi oscillations in defective plasma photonic crystals},\ }\href {https://doi.org/10.1103/k2th-m73q} {\bibfield  {journal} {\bibinfo  {journal} {Phys. Rev. Lett.}\ }\textbf {\bibinfo {volume} {135}},\ \bibinfo {pages} {015101} (\bibinfo {year} {2025})}\BibitemShut {NoStop}%
\bibitem [{\citenamefont {Lehmann}\ and\ \citenamefont {Spatschek}(2016)}]{Lehmann2016}%
  \BibitemOpen
  \bibfield  {author} {\bibinfo {author} {\bibfnamefont {G.}~\bibnamefont {Lehmann}}\ and\ \bibinfo {author} {\bibfnamefont {K.~H.}\ \bibnamefont {Spatschek}},\ }\bibfield  {title} {\bibinfo {title} {Transient plasma photonic crystals for high-power lasers},\ }\href {https://doi.org/10.1103/PhysRevLett.116.225002} {\bibfield  {journal} {\bibinfo  {journal} {Phys. Rev. Lett.}\ }\textbf {\bibinfo {volume} {116}},\ \bibinfo {pages} {225002} (\bibinfo {year} {2016})}\BibitemShut {NoStop}%
\bibitem [{\citenamefont {Wagner}\ \emph {et~al.}(2004)\citenamefont {Wagner}, \citenamefont {Gibson}, \citenamefont {Popmintchev}, \citenamefont {Christov}, \citenamefont {Murnane},\ and\ \citenamefont {Kapteyn}}]{Wagner2004}%
  \BibitemOpen
  \bibfield  {author} {\bibinfo {author} {\bibfnamefont {N.~L.}\ \bibnamefont {Wagner}}, \bibinfo {author} {\bibfnamefont {E.~A.}\ \bibnamefont {Gibson}}, \bibinfo {author} {\bibfnamefont {T.}~\bibnamefont {Popmintchev}}, \bibinfo {author} {\bibfnamefont {I.~P.}\ \bibnamefont {Christov}}, \bibinfo {author} {\bibfnamefont {M.~M.}\ \bibnamefont {Murnane}},\ and\ \bibinfo {author} {\bibfnamefont {H.~C.}\ \bibnamefont {Kapteyn}},\ }\bibfield  {title} {\bibinfo {title} {Self-compression of ultrashort pulses through ionization-induced spatiotemporal reshaping},\ }\href {https://doi.org/10.1103/PhysRevLett.93.173902} {\bibfield  {journal} {\bibinfo  {journal} {Phys. Rev. Lett.}\ }\textbf {\bibinfo {volume} {93}},\ \bibinfo {pages} {173902} (\bibinfo {year} {2004})}\BibitemShut {NoStop}%
\bibitem [{\citenamefont {Pan}\ \emph {et~al.}(2023)\citenamefont {Pan}, \citenamefont {Cohen},\ and\ \citenamefont {Segev}}]{Pan2023}%
  \BibitemOpen
  \bibfield  {author} {\bibinfo {author} {\bibfnamefont {Y.}~\bibnamefont {Pan}}, \bibinfo {author} {\bibfnamefont {M.-I.}\ \bibnamefont {Cohen}},\ and\ \bibinfo {author} {\bibfnamefont {M.}~\bibnamefont {Segev}},\ }\bibfield  {title} {\bibinfo {title} {Superluminal $k$-gap solitons in nonlinear photonic time crystals},\ }\href {https://doi.org/10.1103/PhysRevLett.130.233801} {\bibfield  {journal} {\bibinfo  {journal} {Phys. Rev. Lett.}\ }\textbf {\bibinfo {volume} {130}},\ \bibinfo {pages} {233801} (\bibinfo {year} {2023})}\BibitemShut {NoStop}%
\bibitem [{\citenamefont {Jang}\ \emph {et~al.}(2025)\citenamefont {Jang}, \citenamefont {Oh}, \citenamefont {Kim},\ and\ \citenamefont {Rho}}]{Jang2025}%
  \BibitemOpen
  \bibfield  {author} {\bibinfo {author} {\bibfnamefont {Y.}~\bibnamefont {Jang}}, \bibinfo {author} {\bibfnamefont {B.}~\bibnamefont {Oh}}, \bibinfo {author} {\bibfnamefont {E.}~\bibnamefont {Kim}},\ and\ \bibinfo {author} {\bibfnamefont {J.}~\bibnamefont {Rho}},\ }\bibfield  {title} {\bibinfo {title} {Bidirectional asymmetric frequency conversion in nonlinear phononic crystals},\ }\href {https://doi.org/10.1103/3n97-7kmd} {\bibfield  {journal} {\bibinfo  {journal} {Phys. Rev. Lett.}\ }\textbf {\bibinfo {volume} {135}},\ \bibinfo {pages} {036603} (\bibinfo {year} {2025})}\BibitemShut {NoStop}%
\bibitem [{\citenamefont {Dong}\ \emph {et~al.}(2025{\natexlab{b}})\citenamefont {Dong}, \citenamefont {Zhang}, \citenamefont {He}, \citenamefont {Li},\ and\ \citenamefont {Xu}}]{Dong2025}%
  \BibitemOpen
  \bibfield  {author} {\bibinfo {author} {\bibfnamefont {J.}~\bibnamefont {Dong}}, \bibinfo {author} {\bibfnamefont {S.}~\bibnamefont {Zhang}}, \bibinfo {author} {\bibfnamefont {H.}~\bibnamefont {He}}, \bibinfo {author} {\bibfnamefont {H.}~\bibnamefont {Li}},\ and\ \bibinfo {author} {\bibfnamefont {J.}~\bibnamefont {Xu}},\ }\bibfield  {title} {\bibinfo {title} {Nonuniform wave momentum band gap in biaxial anisotropic photonic time crystals},\ }\href {https://doi.org/10.1103/PhysRevLett.134.063801} {\bibfield  {journal} {\bibinfo  {journal} {Phys. Rev. Lett.}\ }\textbf {\bibinfo {volume} {134}},\ \bibinfo {pages} {063801} (\bibinfo {year} {2025}{\natexlab{b}})}\BibitemShut {NoStop}%
\bibitem [{\citenamefont {Ying}\ \emph {et~al.}(2025)\citenamefont {Ying}, \citenamefont {Liu}, \citenamefont {Zheng}, \citenamefont {Su}, \citenamefont {He}, \citenamefont {Ma}, \citenamefont {Xuan},\ and\ \citenamefont {Zhang}}]{Ying2025}%
  \BibitemOpen
  \bibfield  {author} {\bibinfo {author} {\bibfnamefont {J.}~\bibnamefont {Ying}}, \bibinfo {author} {\bibfnamefont {L.}~\bibnamefont {Liu}}, \bibinfo {author} {\bibfnamefont {L.}~\bibnamefont {Zheng}}, \bibinfo {author} {\bibfnamefont {D.}~\bibnamefont {Su}}, \bibinfo {author} {\bibfnamefont {X.}~\bibnamefont {He}}, \bibinfo {author} {\bibfnamefont {J.}~\bibnamefont {Ma}}, \bibinfo {author} {\bibfnamefont {H.}~\bibnamefont {Xuan}},\ and\ \bibinfo {author} {\bibfnamefont {D.}~\bibnamefont {Zhang}},\ }\bibfield  {title} {\bibinfo {title} {High-energy subcycle electron emission driven by spatiotemporally confined thz fields},\ }\href {https://doi.org/10.1103/wczq-1jv3} {\bibfield  {journal} {\bibinfo  {journal} {Phys. Rev. X}\ }\textbf {\bibinfo {volume} {15}},\ \bibinfo {pages} {021095} (\bibinfo {year} {2025})}\BibitemShut {NoStop}%
\end{thebibliography}
\end{document}